\documentclass[%
reprint,
superscriptaddress,
 amsmath,amssymb,
longbibliography
]{revtex4-2}

\usepackage{aas_macros}
\usepackage{graphicx}
\usepackage{dcolumn}
\usepackage{bm}
\usepackage{hyperref}
\usepackage{color}
\usepackage[mathlines]{lineno}

\usepackage{subfigure}
\usepackage[normalem]{ulem}
\usepackage[export]{adjustbox}
\usepackage{url}
\usepackage{lineno}
\usepackage[flushleft]{threeparttable}
\usepackage{makecell}
\usepackage{multirow}
\usepackage{tabularx}
\usepackage{rotating}

\hypersetup{
    pdfnewwindow=true,    
    colorlinks=true,      
    linkcolor=blue,       
    citecolor=blue,       
    filecolor=blue,       
    urlcolor=blue         
}

\def\orcid#1{\href{https://orcid.org/#1}{\includegraphics[keepaspectratio,width=1.1em]{Fig/orcid.png}}}
\usepackage{natbib}

\begin{document}
\title{Measurements of All-Particle Energy Spectrum and Mean Logarithmic Mass of Cosmic Rays from 0.3 to 30 PeV with LHAASO-KM2A}
 
\author{Zhen Cao}
\affiliation{Key Laboratory of Particle Astrophysics \& Experimental Physics Division \& Computing Center, Institute of High Energy Physics, Chinese Academy of Sciences, 100049 Beijing, China}
\affiliation{University of Chinese Academy of Sciences, 100049 Beijing, China}
\affiliation{Tianfu Cosmic Ray Research Center, 610000 Chengdu, Sichuan,  China}
 
\author{F. Aharonian}
\affiliation{Dublin Institute for Advanced Studies, 31 Fitzwilliam Place, 2 Dublin, Ireland }
\affiliation{Max-Planck-Institut for Nuclear Physics, P.O. Box 103980, 69029  Heidelberg, Germany}
 
\author{Axikegu}
\affiliation{School of Physical Science and Technology \&  School of Information Science and Technology, Southwest Jiaotong University, 610031 Chengdu, Sichuan, China}
 
\author{Y.X. Bai}
\affiliation{Key Laboratory of Particle Astrophysics \& Experimental Physics Division \& Computing Center, Institute of High Energy Physics, Chinese Academy of Sciences, 100049 Beijing, China}
\affiliation{Tianfu Cosmic Ray Research Center, 610000 Chengdu, Sichuan,  China}
 
\author{Y.W. Bao}
\affiliation{School of Astronomy and Space Science, Nanjing University, 210023 Nanjing, Jiangsu, China}
 
\author{D. Bastieri}
\affiliation{Center for Astrophysics, Guangzhou University, 510006 Guangzhou, Guangdong, China}
 
\author{X.J. Bi}
\affiliation{Key Laboratory of Particle Astrophysics \& Experimental Physics Division \& Computing Center, Institute of High Energy Physics, Chinese Academy of Sciences, 100049 Beijing, China}
\affiliation{University of Chinese Academy of Sciences, 100049 Beijing, China}
\affiliation{Tianfu Cosmic Ray Research Center, 610000 Chengdu, Sichuan,  China}
 
\author{Y.J. Bi}
\affiliation{Key Laboratory of Particle Astrophysics \& Experimental Physics Division \& Computing Center, Institute of High Energy Physics, Chinese Academy of Sciences, 100049 Beijing, China}
\affiliation{Tianfu Cosmic Ray Research Center, 610000 Chengdu, Sichuan,  China}
 
\author{W. Bian}
\affiliation{Tsung-Dao Lee Institute \& School of Physics and Astronomy, Shanghai Jiao Tong University, 200240 Shanghai, China}
 
\author{A.V. Bukevich}
\affiliation{Institute for Nuclear Research of Russian Academy of Sciences, 117312 Moscow, Russia}
 
% ...
\collaboration{The LHAASO Collaboration}

\email{hyzhang@ihep.ac.cn(H.Y. Zhang)}
\email{hhh@ihep.ac.cn(H.H. He)}
\email{fengcf@sdu.edu.cn(C.F. Feng)}

\date{\today}

\begin{abstract}

We present the measurements of all-particle energy spectrum and mean logarithmic mass of cosmic rays in the energy range of 0.3$-$30 PeV using data collected from LHAASO-KM2A between September 2021 and December 2022, which is based on a nearly composition-independent energy reconstruction method, achieving unprecedented accuracy. Our analysis reveals the position of the knee at $3.67 \pm 0.05 \pm 0.15$ PeV. Below the knee, the spectral index is found to be $-2.7413 \pm 0.0004 \pm 0.0050$, while above the knee, it is $-3.128 \pm 0.005 \pm 0.027$, with the sharpness of the transition measured with a statistical error of 2\%. The mean logarithmic mass of cosmic rays is almost heavier than helium in the whole measured energy range. It decreases from 1.7 at 0.3 PeV to 1.3 at 3 PeV, representing a 24\% decline following a power law with an index of $-0.1200 \pm 0.0003 \pm 0.0341$. This is equivalent to an increase in abundance of light components. Above the knee, the mean logarithmic mass exhibits a power law trend towards heavier components, which is reversal to the behavior observed in the all-particle energy spectrum. Additionally, the knee position and the change in power-law index are approximately the same. These findings suggest that the knee observed in the all-particle spectrum corresponds to the  knee of the light component, rather than the medium-heavy components.

\end{abstract}

\maketitle


\textit{Introduction.---}The energy spectrum of cosmic rays spans a wide range from $10^{9}$ to $10^{20}$ eV, roughly following a power law of ${\rm{d}}N/{\rm{d}}E \propto E^{\gamma}$, where $\gamma$ is the differential spectral index. Features in the spectrum correspond to changes in $\gamma$. At approximately 4 PeV, the differential spectral index changes from $\gamma \approx -2.7$ at low energies to $\gamma\approx -3.1$. This structure is referred to as the "knee"~\cite{kulikov1959size}. The knee as the beginning of the transition of high energy cosmic rays from galactic to extra-galactic origin is still controversial. The energy spectrum and chemical composition of cosmic rays can provide crucial clues regarding the origin, acceleration, and propagation mechanisms of the most energetic particles in the galaxy \cite{HORANDEL2004241,Piazzoli_2022}. Understanding the origin of the
knee in terms of the energy spectrum is often considered fundamental for determining the origin of cosmic rays. Space-borne experiments have now extended the direct measurement of the energy spectrum of proton and helium nuclei to 100 TeV ~\citep{DAMPE1,CREAMIII}. Because of the rapidly decreasing flux with increasing energy, measuring cosmic rays beyond 100 TeV by space-borne experiments is challenging. The data for the spectrum above a few hundred TeV are produced by ground-based large air shower arrays. The knee structure at few PeV in all-particle energy spectrum of cosmic rays has been observed by many experiments~\cite{TibetIII2008,ICETOP40,PhysRevD882013,IceCubeandTop,ICETOP2020,KASCANDAPP2005,KASCADEICRC2017,tunka201298,CASAMIA1999,CASA1999291}. Ground-based indirect experiment is the only way to measure cosmic rays with energies in the knee region, while the main problem with these experiments arises from the entanglement of primary energy, composition and the high-energy hadronic interaction models used to interpret air showers. The chemical composition and high-energy hadronic interaction models dependent systemic errors in flux were a few tens of percent. Recently, some high-energy hadronic interaction models were updated based on new data from the Large Hadron Collider (LHC) at TeV energies~\cite{QGSJET,PRCEPOS}. Although a single-composition energy spectrum is the most effective way for solving this problem, identifying the composition is challenging. Thus, the problem of composition dependence on energy reconstruction must be solved to measure cosmic ray energy spectra accurately.

The Large High Altitude Air Shower Observatory (LHAASO) is a ground-based air shower observatory at 4410 m above sea level in Daocheng, China \citep{CzhenCPC,HHHRDTM,Ma2022}. The high altitude of the observatory enables LHAASO to measure showers with energies in the knee region close to their maxima with minimum fluctuations. It consists of three detector arrays: a kilometer square array (KM2A), a water Cherenkov detector array (WCDA), and a wide-field-of-view Cherenkov or fluorescence telescope array (WFCTA). More details about the experimental setup can be found in \cite{HHHRDTM}. The array used in this measurement is the KM2A that measures electromagnetic particles (electron, positron and gamma) and muons of air showers with high precision \cite{CPCCrab} by the electromagnetic particle detectors (EDs) and muon detectors (MDs), respectively. The MD threshold for $\mu^{\pm}$ is about 1 GeV \cite{LicongPRD}. The detector layout is shown in Fig. S1 of the Supplemental Material \cite{SM}. The number of electromagnetic particles ($N_{\rm{e}}$) and muons ($N_{\rm{\mu}}$) is the deposited charge by electromagnetic particles and muons expressed in units of minimum-ionizing particles. Basing on the Matthews-Heitler model \cite{Mat-HeitlerApp2005}, a new parameter, derived by summing $N_{\rm{e}}$ and $N_{\rm{\mu}}$, is used to reconstruct primary energy in a calorimetric way. This approach is insensitive to primary composition and hadronic interaction models thus is crucial for measuring accurately the energy spectrum.


\textit{Data.---}Data collected by the LHAASO-KM2A from September 2021 to December 2022 are utilized for this analysis. A maximum variation of approximately 12 hPa was observed in the atmospheric pressure, resulting in a change of about 4.4\% in the muon spectra, as illustrated in Figs. S2 and S3 of the Supplemental Material \cite{SM}. The constant intensity cut (CIC) method ~\cite{PhysRevLett622CIC, APEL201725} is employed to correct for the atmospheric pressure effect. Subsequently, the discrepancy in the number of muons under different atmospheric pressures is reduced to less than 0.6\%. 

The CORSIKA (v77410) air shower generator \cite{Corsikaguide} with the high-energy interaction models QGSJETII-04 \cite{QGSJET}, EPOS-LHC \cite{PRCEPOS} and SIBYLL-2.3d \cite{PRDSIBYLL}, is used to generate air-shower events. For the low-energy interactions the FLUKA \cite{Fluka201510} code is used. The detector response simulation is based on the Geant4 package \cite{Geant4}, named G4KM2A \cite{ChenSZ2017NEDT,CPCCrab}. The simulated detector response agrees with the response of a real detector \cite{CPCCrab}. The Monte Carlo simulation is described in detail in the Supplemental Material \cite{SM}.

\textit{Event selection.---}High-energy electromagnetic particles near shower cores can punch through the MD overburden soil, which can  pollute the measurement of muons~\cite{ZhanghyPRD,ZUO2015}. In order to mitigate this effect, $N_{\rm{\mu}}$ within a range of $40-200$ m from the shower core is selected for analysis. To ensure consistency in the analysis, $N_{\rm{e}}$ within the same range of $40-200$ m from the shower core is also taken into account. The LHAASO is located at an altitude of 4,410 m above sea level with an vertical atmospheric depth of 600 g/cm$^{2}$, which corresponds to the shower maxima of cosmic rays with energies in the knee region. To capture cosmic ray showers near their maxima within an energy range of 2 orders of magnitude, events are selected with zenith angles ranging from 10$^\circ$ to 30$^\circ$, which corresponds to an atmosphere depth of 610 and 690 g/cm$^2$, respectively. It is important to note that cosmic rays of different energies reach their shower maxima at varying atmospheric depths. Further information regarding the event selection process can be found in the Supplemental Material \cite{SM}. Above 0.3 PeV, all five mass groups of cosmic ray components reach full detection efficiency, as shown in Fig. S4 of the Supplemental Material \cite{SM}. Therefore, the geometric aperture after considering core location and zenith angle selection can be derived using the following equation:
\begin{equation}
    A_{\rm{eff}}=S\int^{\theta_{2}}_{\theta_{1}}\rm sin\theta \rm cos\theta d\theta\int^{2\pi}_{0}\rm d\varphi ,
\end{equation}
Here, $S$ represents the area of the $320-420$ m ring from the center of the KM2A, [$\theta_{1}$, $\theta_{2}$] is the observed zenith range with $\theta_{1}$=10$^{\circ}$ and $\theta_{2}$=30$^{\circ}$. The geometric aperture $A_{\rm{eff}}$ is determined to be 0.16 km$^2$sr. In this Letter, we present the all-particle energy spectrum and the mean logarithmic mass of cosmic rays based on a dataset of $7\times 10^7$ events, covering an energy range from 0.3 to 30 PeV.

\textit{Energy reconstruction.---}Basing on the Matthews-Heitler model \cite{Mat-HeitlerApp2005}, a nearly composition-independent parameter $N_{{\rm{e\mu}}}$ is adopted to reconstruct the primary energy of cosmic rays in a calorimetric way, which is defined as follows:
\begin{equation}
N_{\rm{e\mu}}=N_{\rm{e}}+2.8N_{{\rm{\mu}}}
\end{equation}
where the calculation of $N_{\rm{e}}$ (or $N_{\rm{\mu}}$) considers only the the EDs (or MDs) within $40-200$ m from the shower axis~\cite{ZhanghyPRD}. In this Letter, the energy reconstruction formula differs slightly from the previous article~\cite{ZhanghyPRD} where quasivertical events within one energy decade were selected. Instead, the zenith angle range of 10$^\circ-30^\circ$ is chosen for cosmic rays with energy covering 2 orders of magnitude. The energy reconstruction formula used is $\log_{10}(E/{\rm{GeV}})=p_0+p_1\cdot\log_{10}(N_{\rm{e\mu}})$, details can be found in the Supplemental Material \cite{SM}. Above a few hundred TeV, the correlation between $N_{\rm{e\mu}}$ and the primary energy becomes nearly independent of the components as demonstrated in Fig. S5 using the QGSJETII-04 model and the Gaisser H3a model as examples in the Supplemental Material \cite{SM}. This suggests that the composition of the cosmic rays does not significantly impact the accuracy of the energy reconstruction based on $N_{\rm{e\mu}}$. The differences of parameters $p_0$ and $p_1$ in energy reconstruction formulas for different composition models (such as Gaisser H3a \cite{Gaisser2013}, Horandel \cite{HORANDEL2003193}, GST \cite{STANEV201442} and GSF \cite{ICRCGSF2017}) are very small, as shown in Table S1 of the Supplemental Material \cite{SM}. Therefore, the energy reconstruction formulas for each high-energy hadronic interaction models are based on the average parameters of four composition models, as shown in Table S2 of the Supplemental Material \cite{SM}. Different hadronic interaction models result in energy reconstruction formulas with little difference, so the average one is used to reconstruct cosmic ray energy, which is $\log_{10}(E/{\rm{GeV}})=2.791+0.993\cdot\log_{10}(N_{\rm{e\mu}})$.


\textit{All-particle energy spectrum of cosmic rays.---}The core location, direction, and energy of a shower are reconstructed, and the binned flux can be calculated using the following formula
\begin{equation}
J(E)=\frac{{\Delta}N(E)}{{\Delta}E\cdot A_{\rm{eff}}\cdot T} ,
\label{flux-cal}
\end{equation}
where ${\Delta}N$ represents the number of events per energy bin in a given time interval $T$, and $A_{\rm{eff}}$ represents the geometric aperture. To validate this method, simulation data are analyzed in the same way, and the results are presented in Fig. S6 of the Supplemental Material \cite{SM}, which confirm that our method can reproduce all the input spectra with high accuracy. 

\begin{figure}[htpb]
\centering
\includegraphics[width=0.45\textwidth]{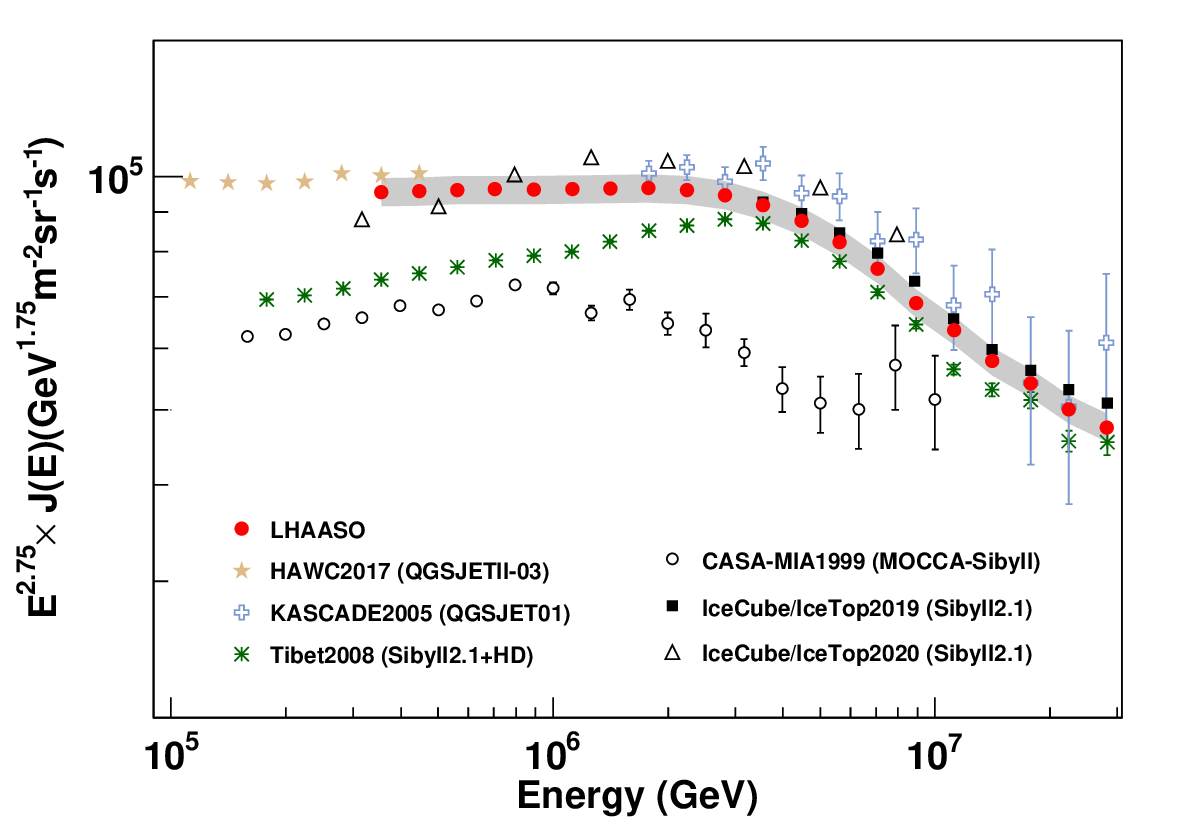}
\caption{\fontsize{8pt}{\baselineskip}\selectfont{The LHAASO-KM2A all-particle energy spectrum flux multiplied by $E^{2.75}$ as a function of energy. The gray shadow band represents the estimated systematic uncertainties. The results of the HAWC~\cite{PhysRevD.96.122001}, CASA-MIA \cite{CASA1999291}, Tibet-III \cite{TibetIII2008}, KASCADE \cite{KASCANDAPP2005}, IceCube/IceTop \cite{IceCubeandTop,ICETOP2020} are also plotted for comparison.  The error bars  represent the statistical uncertainties.}}
\label{allspectrum}
\end{figure}

The all-particle energy spectrum of primary cosmic rays is calculated following Eq. (\ref{flux-cal}) within the energy range of 0.3 to 30 PeV, as shown in Fig.\ref{allspectrum} from which the knee structure is clearly visible.

The systematic error of the spectrum is considered as follows. The single particle calibration uncertainty of one ED and MD is less than $2\%$ \cite{lvPRD} and 0.5\% \cite{ZUO2015}, respectively. At 300 TeV, the average number of triggered EDs and MDs is approximately 200 and 50, respectively. Therefore, the measurement uncertainty of $N_{\rm{e}}$ and $N_{\rm{\mu}}$ are both less than 0.15\%, which contribute a negligible uncertainty on all-particle energy spectrum. The impact of local variations in air pressure on the results of this study is estimated to be approximately $\pm 3\%$. The systematic uncertainty observed between different composition models, such as Gaisser H3a, Horandel, GST, and GSF models, are found to be less than $\pm 1.5\%$. The influence of QGSJETII-04, EPOS-LHC, and SIBYLL-2.3d high-energy hadronic interaction models on the flux systematic uncertainty is approximately $\pm 2.5\%$ as shown in Fig. S9 of the Supplemental Material \cite{SM}. The shadow band in Fig. \ref{allspectrum} represents the systematic uncertainties associated with air pressure, composition models, and different high-energy hadronic interaction models. A comparison of these systematic errors can be found in Table S3 of the Supplemental Material \cite{SM}. The values of the all-particle energy spectrum, along with the corresponding statistical and combined systematic uncertainties (excluding the uncertainty from hadronic interaction models), can be found in Table S5 of the Supplemental Material \cite{SM}.

The AS$\gamma$  experiment shows that the difference between the pure proton and pure iron primary models becomes larger in the lower energy region, where the difference in the intensities between the two models exceeds a factor of 3 at $10^{14}$ eV~\cite{TibetIII2008}. The IceTop-73 results show that the difference of energy spectrum intensities between assuming pure proton and pure iron models exceeds a factor of 1.5 at 2 PeV \cite{PRD2013IceTop}. Of course, the extreme cases of pure proton and pure iron do not reflect reality. In this work, the flux difference between the pure proton and pure iron models is approximately 12\%, which is better by more than a factor of 10 than that of previous experiments and thus allows for a more robust and reliable estimation of the primary energy of cosmic rays.

\begin{figure}
\centering
\includegraphics[width=0.45\textwidth]{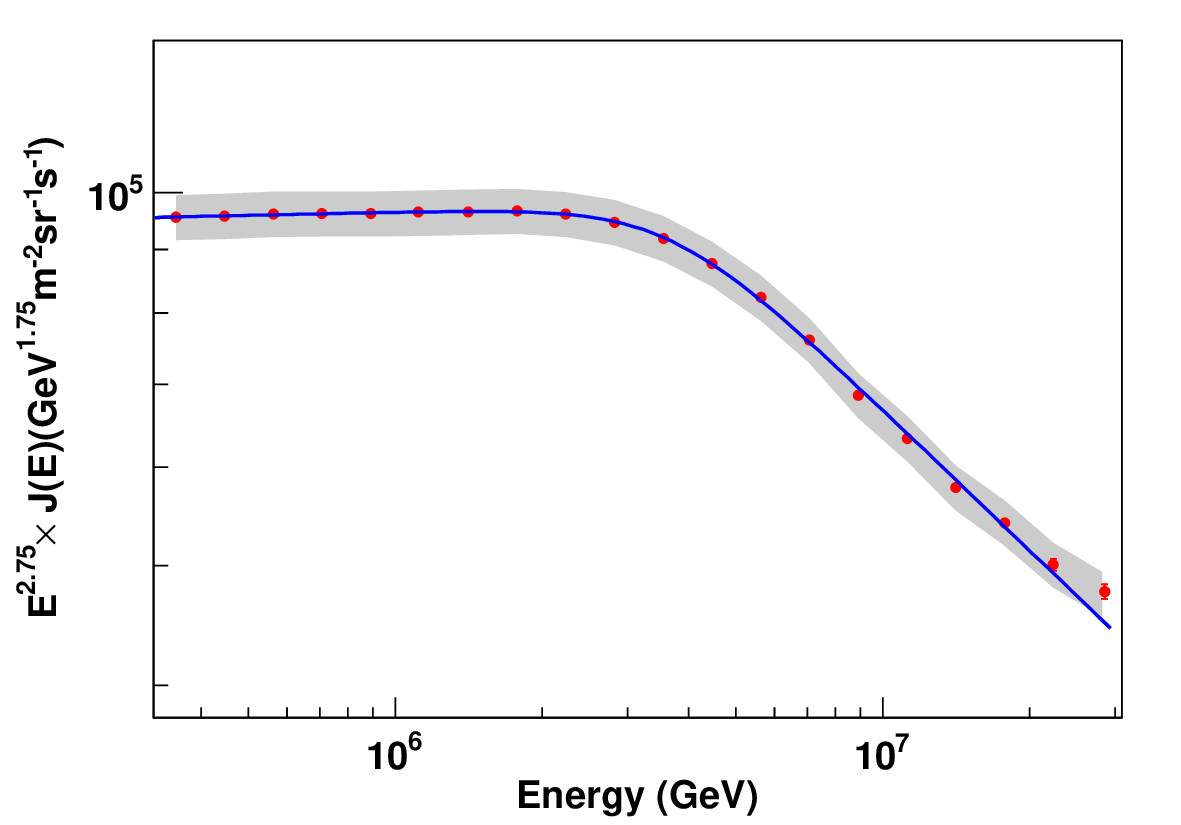}
\includegraphics[width=0.45\textwidth]{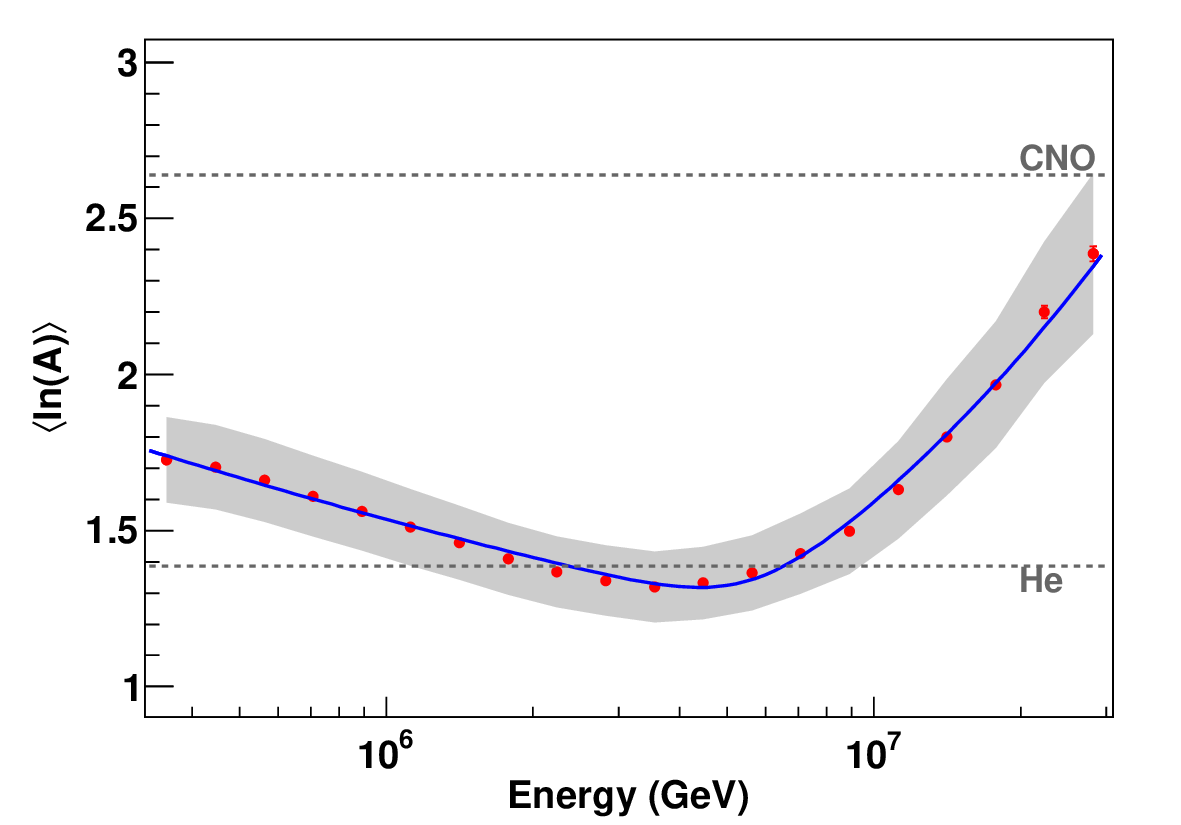}
\caption{\fontsize{8pt}{\baselineskip}\selectfont{Upper: The all-particle energy spectrum flux multiplied by $E^{2.75}$ as a function of energy. Bottom: The $\langle\ln (A)\rangle$ of cosmic rays as a function of energy. The error bars show the statistical uncertainties and the gray shadow band represents the estimated combined systematic uncertainties. The blue solid curves indicate the fit of Equation (\ref{fitflux}) to the data.  }}
\label{Fitall}
\end{figure}

Figure \ref{Fitall} shows the spectrum fitted with the following formula~\citep{SVT2000eas,KASCANDAPP2005}
\begin{equation}
J(E)=\Phi_0\cdot(E)^{\gamma_1}\left(1+\left( \frac{E}{E_{\rm{b}}}\right)^{s} \right)^{(\gamma_2-\gamma_1)/s} 
\label{fitflux}
\end{equation}
where $E_{\rm{b}}$ corresponds to the knee position, $\gamma_1$ and $\gamma_2$ are spectral asymptotic slopes before and after knee, $s$ is the sharpness parameter of the knee. According to our fitting results, the knee position is at $3.67 \pm 0.05 \pm 0.15$ PeV, where the first term is statistical error and the second is systematic error. The spectral indices are $-2.7413 \pm 0.0004 \pm 0.0050$ and $-3.128 \pm 0.005 \pm 0.027$ before and after the knee, respectively. The energy spectral index changes $0.387\pm 0.005 \pm 0.027$ before and after the knee, with the sharpness of the transition measured at $4.2\pm0.1\pm0.5$. This study represents one of the most precise measurements of the all-particle energy spectrum to date. The numerical values for different high-energy hadronic interaction models are summarized in Table S4 of the Supplemental Material \cite{SM}. 

\textit{Mean logarithmic mass of cosmic rays.---}According to the Matthews-Heitler model \cite{Mat-HeitlerApp2005}, the muon content of a cosmic ray shower is dependent on its composition and energy as follows:
\begin{equation}
N_{\rm{\mu}}{\propto}A\cdot\left(\frac{E}{A\cdot\varepsilon_{c}}\right) ^\beta 
\label{EqNu}
\end{equation}
where $A$ represents the mass of the cosmic ray, with a value of 1 for a proton and 56 for iron. Additionally, $\varepsilon_{c}$ refers to the critical energy at shower maximum~\citep{PhysRevD.66.033011}. According to Eq. (\ref{EqNu}), given the energy, the relationship between the mean of the logarithmic muon content [$\langle\ln (N_{\rm{\mu}})\rangle$] and the mean logarithmic mass of cosmic rays [$\langle\ln (A)\rangle$] can be determined 
\begin{equation}
\langle \ln (N_{\rm{\mu}})\rangle=x_0+x_1\cdot\langle\ln (A)\rangle. 
\label{EqlnA}
\end{equation}
The parameters $x_0$ indicate the proton $\langle \ln (N_{\rm{\mu}})\rangle_{\rm H}$ and $x_1$ indicate $1-\beta$ \cite{HPDAPP2018}. These parameters are obtained through fitting simulation data. To reduce the statistical fluctuation of the simulation data, the energy interval of the simulation data is set to 0.2, as shown in the left plot of Fig. S7 of Supplemental Material \cite{SM}. These parameters, $x_0$ and $x_1$, exhibit variations with energy within the energy range of $\log_{10}(E_{\rm{rec}}/\rm GeV)=$5.5 and 7.5, as shown in Fig. S7 of Supplemental Material \cite{SM} using the QGSJETII-04 model as an example. By fitting the relationship between $x_0$ or $x_1$ and cosmic ray energy, we can interpolate to obtain the corresponding values of $x_0$ or $x_1$ within a 0.1 energy interval. The validation of this method, using four assumed composition models, is illustrated in Fig. S8 of the Supplemental Material \cite{SM}. These figures confirm that our method can reproduce all the input composition models with high accuracy. 

\begin{figure}
\centering
\includegraphics[width=0.45\textwidth]{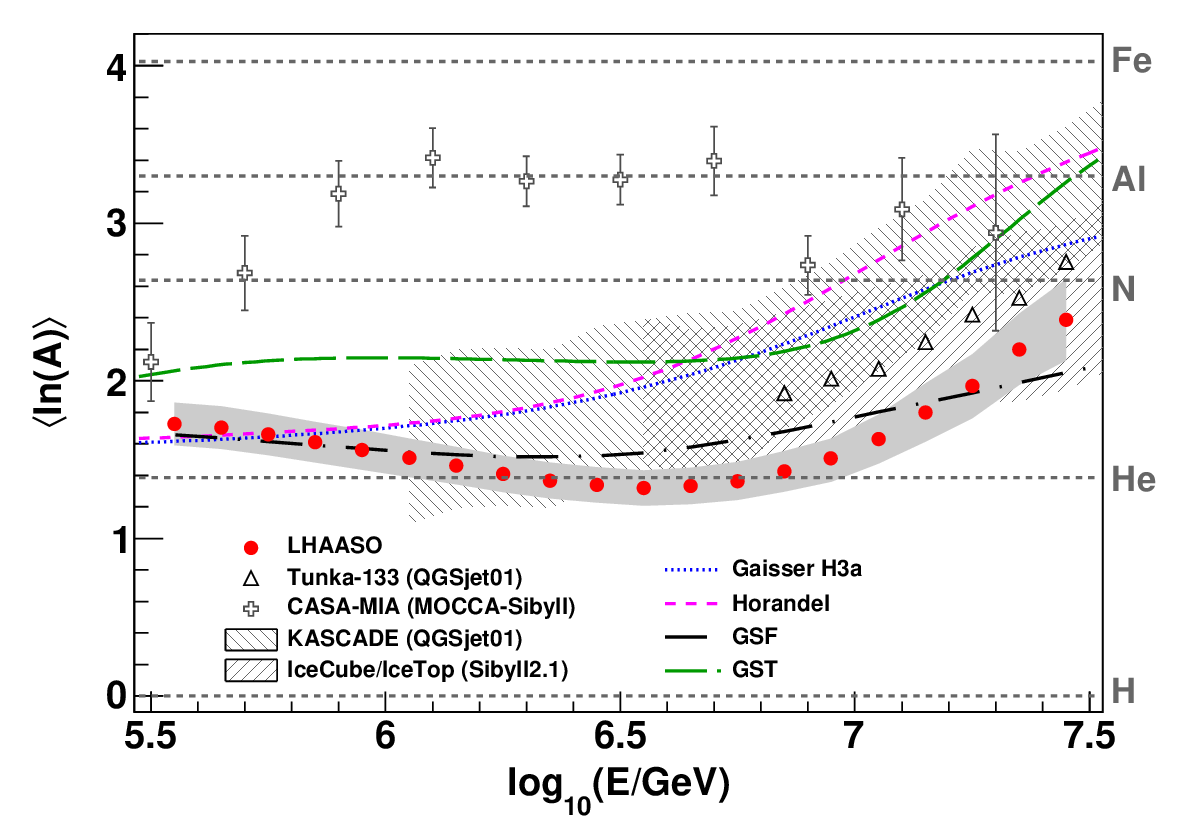}
\caption{\fontsize{8pt}{\baselineskip}\selectfont{ The $\left\langle\ln (A)\right\rangle$ (red dots) from experimental data as a function of cosmic rays energy. The error bars show the statistical uncertainties. The gray shadow band represents the combined systematic uncertainties. The results of CASA-MIA \cite{CASAMIA1999} and Tunka-133 \cite{Tunka133NIMA2012} are also plotted for comparison. The shadow bands represent the systematic uncertainty for KASCADE~\cite{KASCANDAPP2005,KAMPERT2012660} and IceCube/IceTop~\cite{IceCubeandTop} experiment, respectively. The curves are the $\left\langle\ln (A)\right\rangle$ which calculated according the cosmic rays composition models. }}
\label{lnAall}
\end{figure}

Using the above method, LHAASO measures the $\langle\ln (A)\rangle$ in each cosmic ray energy bin from 0.3 to 30 PeV, and these measurements are shown in Fig. \ref{lnAall}, along with results from other experiments. The distribution of $\langle\ln (A)\rangle$ predicted by Gaisser H3a, Horandel, GST, and GSF models are also depicted in Fig. \ref{lnAall}. At low energy region around 300 TeV, our results are consistent with the predictions of the Horandel, Gaisser H3a models. However, as the energy increases, our results deviate from the predictions of the Horandel or Gaisser H3a models, indicating that these models are not able to fully explain our observations.

The systematic error of the $\langle\ln (A)\rangle$ is considered as follows. The impact of local variations in air pressure on $\langle\ln (A)\rangle$ is approximately $\pm 4\%$. The differences of $\langle\ln (A)\rangle$   between several cosmic ray composition models, such as Gaisser H3a, Horandel, GST, and GSF models, are found to be less than $\pm 3\%$. The distribution of $\left\langle\ln (A)\right\rangle$ for the QGSJETII-04, EPOS-LHC, and SIBYLL-2.3d models exhibits a very similar shape, as shown in Fig. S9 of the Supplemental Material \cite{SM}, and the maximum difference between them is less than $\pm 6\%$. The shadow band in Fig. \ref{lnAall} represents the systematic uncertainties associated with air pressure, composition models, high-energy hadronic interaction models and the parameters $x_0$ and $x_1$ obtained by fitting the simulation data.  The detail values of these systematic errors can be seen in Table S3 of the Supplemental Material \cite{SM}. The values of $\langle\ln (A)\rangle$, along with their corresponding statistical and combined systematic uncertainties (excluding the uncertainty from hadronic interaction models), are included in Table S6 of the Supplemental Material \cite{SM}.

The results of other experiments, whose energy regions overlap with this work, are also plotted for comparison. While measurements from other experiments show a gradual increase in the mean logarithmic mass, our measurement reveals a decrease followed by an increase,  with the sharpness of the transition measured at $5.5\pm 0.3\pm 0.9$. The systematic uncertainty of our measurement results is $\pm 9\%$, indicated by the gray shadow band in Fig. \ref{lnAall}. In comparison, both the IceCube/IceTop experimental~\cite{IceCubeandTop} and KASCADE~\cite{KAMPERT2012660} measurement results exhibit a systematic uncertainty of approximately $\pm 25\%$, represented by the shadow bands in Fig. \ref{lnAall}. This study is the most precise measurement of $\langle\ln (A)\rangle$ in this energy region to date. The $\langle\ln (A)\rangle$ from different theories has been studied in \cite{Hadron-EPJ2019} for the energy region from  $10^{15}$ to $10^{17}$ eV. Results on $\langle\ln (A)\rangle$ from the muon content $N_{\rm{\mu}}$ and those from the depth of shower maximum $X_{\rm{max}}$ show the same tendency toward heavier $\langle\ln (A)\rangle$ with energy beyond several PeV. Our results agree with this expectation in this energy region, where the $\langle\ln (A)\rangle$ becomes heavier with energy beyond several PeV. 

The fitting results of the all-particle energy spectrum and the $\langle\ln (A)\rangle$ of the cosmic ray using Eq. (\ref{fitflux}) are shown in Fig. \ref{Fitall}.  We find an interesting feature of the $\langle\ln (A)\rangle$ of cosmic rays: it decreases by 24\% between 0.3 and 3 PeV and becomes heavier. The $\langle\ln (A)\rangle$ is closest to He [$\ln(A)=1.39$], and lighter than CNO [$\ln(A)=2.64$], suggesting that the first cutoff of the all-particle energy spectrum is due to light components, instead of the medium-heavy components. Another interesting feature is that the mean logarithmic mass exhibits a variation with energy, which is reversal to the behavior observed in the all-particle energy spectrum. The fitting results of $\langle\ln (A)\rangle$ has a spectral index of $\gamma_{1}=-0.1200\pm 0.0003\pm 0.0341$ before the energy where the composition of cosmic rays shows a clear change from helium on average toward the heavier CNO on average. The corresponding change in the spectral index is $\Delta \gamma=0.497\pm 0.008\pm 0.039$, occurring at an energy position of $5.36 \pm 0.08\pm 0.46$ PeV. In our measurements, we obtained an all-particle energy spectrum index of $-2.74$ before the knee, while the $\langle\ln (A)\rangle$ distribution exhibits a decrease with a spectral index of $-0.12$. This could be a hint that the energy spectrum index of the light component is $[-2.74 - (-0.12)] = -2.62$ before the knee. The fit results for EPOS-LHC, QGSJETII-04 and SIBYLL-2.3d high-energy hadronic interaction models of $\langle\ln (A)\rangle$ can be found in Table S7 of the Supplemental Material \cite{SM}.


\textit{Conclusions.---}In this Letter, the data collected by the LHAASO-KM2A array from September 2021 to December 2022 is used to measure all-particle energy spectrum of cosmic rays from 0.3 to 30 PeV. A composite variable $N_{\rm{e\mu}}$, combining the number of electromagnetic particles and muons, is weakly dependent on the composition mass of cosmic rays and is used to reconstruct the shower energy. The knee of the primary energy spectrum is observed around $3.67 \pm 0.05 \pm 0.15$ PeV with a change of index $\approx 0.387\pm 0.005 \pm 0.027$. The combined systematic uncertainties caused by air pressure, composition models, and high-energy hadronic interaction models are approximately $\pm 4.5\%$. The difference between the pure proton and pure iron assumption models is approximately 12\%.  

The $\left\langle \ln (N_{{\mu}})\right\rangle$ in cosmic rays air showers with energies from 0.3 to 30 PeV is measured by KM2A detectors. The relationship between $\left\langle \ln (N_{\rm{\mu}})\right\rangle$ and $\left\langle\ln (A)\right\rangle$ of cosmic rays is studied using the simulation data with EPOS-LHC, QGSJETII-04 and SIBYLL-2.3d as high-energy hadronic interaction models. The $\left\langle\ln (A)\right\rangle$ exhibits a variation with energy, which is reversal to the behavior observed in the all-particle energy spectrum. The derived $\langle\ln (A)\rangle$ becomes heavier after several PeV. The distribution of $\langle\ln (A)\rangle$ suggests that the first cutoff of the all-particle energy spectrum is light components, not the medium-heavy components. The combined systematic uncertainties, which include air pressure, composition models, the uncertainty of the parameters $x_0$ and $x_1$ obtained by fitting and high-energy hadronic interaction models are approximately $\pm 9\%$. 


We would like to thank all staff members who work at the LHAASO site above 4400 meters above sea level year round to maintain the detector and keep the water recycling system, electricity power supply and other components of the experiment operating smoothly. We are grateful to Chengdu Management Committee of Tianfu New Area for the constant financial support for research with LHAASO data. This research work is also supported by the following grants: The National Key R \& D program of China under Grants No. 2018YFA0404201, No. 2018YFA0404202 and No. 2018YFA0404203, No. 2018YFA0404204, by the National Natural Science Foundation of China(NSFC No.12022502, No.12205314, No. 12105301, No. 12261160362, No.12105294, No.U1931201, No.12175121, No.12275280). We are grateful to the Institute of Plateau Meteorology, CMA Chengdu to maintain meteorological data, and in Thailand by the National Science and Technology Development Agency (NSTDA) and the National Research Council of Thailand (NRCT) under the High-Potential Research Team Grant Program (N42A650868). We are particularly grateful to Dr. L. Q. Yin for the cross-check.

\bibliography{reference}

\end{document}